\RequirePackage{ifpdf}
\ifpdf 
\documentclass[pdftex]{sigma}
\else
\documentclass{sigma}
\fi

\begin{document}
\renewcommand{\atop}[2]{\genfrac{}{}{0pt}{}{#1}{#2}}

\allowdisplaybreaks

\renewcommand{\PaperNumber}{033}

\FirstPageHeading

\ShortArticleName{An Exactly Solvable Spin Chain Related to Hahn Polynomials}

\ArticleName{An Exactly Solvable Spin Chain Related\\ to Hahn Polynomials}

\Author{Neli I. STOILOVA~$^{\dag\ddag}$ and Joris VAN DER JEUGT~$^\ddag$}
\AuthorNameForHeading{N.I.~Stoilova and J.~Van der Jeugt}

\Address{$^\dag$~Institute for Nuclear Research and Nuclear Energy, Boul. Tsarigradsko Chaussee 72,\\
\hphantom{$^\dag$}~1784 Sofia, Bulgaria}

\Address{$^\ddag$~Department of Applied Mathematics and Computer Science, Ghent University,\\
\hphantom{$^\ddag$}~Krijgslaan 281-S9, B-9000 Gent, Belgium}
\EmailD{\href{mailto:Neli.Stoilova@UGent.be}{Neli.Stoilova@UGent.be}, \href{mailto:Joris.VanderJeugt@UGent.be}{Joris.VanderJeugt@UGent.be}}

\ArticleDates{Received January 25, 2011, in f\/inal form March 22, 2011;  Published online March 29, 2011}

\Abstract{We study a linear spin chain which was originally introduced by Shi et al.~[{\em Phys. Rev.~A} {\bf 71} (2005), 032309, 5~pages],
for which the coupling strength contains a parameter $\alpha$
and depends on the parity of the chain site.
Extending the model by a second parameter~$\beta$, it is shown that the single fermion eigenstates of the Hamiltonian can be
computed in explicit form.
The components of these eigenvectors turn out to be Hahn polynomials with parameters $(\alpha,\beta)$ and $(\alpha+1,\beta-1)$.
The construction of the eigenvectors relies on two new dif\/ference equations for Hahn polynomials.
The explicit knowledge of the eigenstates leads to a closed form expression for the correlation function of the spin chain.
We also discuss some aspects of a $q$-extension of this model.}


\Classification{81P45; 33C45}

\section{Introduction}

Consider a linear chain of $N+1$ interacting fermions described by the Hamiltonian
\begin{gather}
\hat H= \sum_{k=0}^{N-1} J_k\big(a_k^\dagger a_{k+1}+a_{k+1}^\dagger a_k\big) .
\label{Ham2}
\end{gather}
The lattice fermions $\{a_{k}, a_{k}^{\dagger}\, |\, k = 0, 1, \ldots, N\}$ obey the common anticommutation relations,
and~$J_k$ expresses the coupling strength between sites~$k$ and~$k+1$.
The Hamiltonian~\eqref{Ham2} describes a~system of $N+1$ fermions on a chain with
nearest-neighbour interaction (hopping between adjacent sites of the chain) subject to a zero background
magnetic f\/ield.

Hamiltonians of the type~\eqref{Ham2} appear in various contexts.
In particular, spin chains of this type are popular as channels for short distance quantum communication, and were introduced by Bose~\cite{Bose2003,Bose2005,Bose2007}.
The system then originates from a linear qubit chain with nearest neighbour interaction described by a Heisenberg $XY$ Hamiltonian,
and is being mapped into~\eqref{Ham2} by a~Jordan--Wigner transformation~\cite{Lieb1968,Jordan1928}.
In such models, the communication is achieved by state dynamical evolution in the spin chain, which does not require
any on/of\/f switches of the interactions between the spins, nor any modulation of external f\/ields.
Many articles dealing with such spin chains in the context of Bose's scheme focus on perfect transmission (or perfect state transfer)
in these chains~\cite{Christandl2004,Albanese2004,Christandl2005,Yung2005,Karbach2005,Kay2010}.

By far the most elegant and simplest scheme to realize perfect state transfer (over an arbitrary long chain) was
proposed by Christandl et al.~\cite{Christandl2004,Albanese2004,Christandl2005}.
Their choice of the modulation of the coupling strengths is given by:
\begin{gather}
J_k= J_{N-k-1}=\sqrt{(k+1)(N-k)}, \qquad k=0,1,\ldots,N-1.
\label{J}
\end{gather}
The simplicity of Christandl's model follows from the following observation.
Consider f\/irst the single-fermion states of the system: in a single-fermion basis, the Hamiltonian $\hat H$ takes the matrix form
\begin{gather}
M=\left(
\begin{array}{ccccc}
0 & J_0& 0 & \cdots & 0 \\
J_0 & 0 & J_1 & \cdots & 0\\
0 & J_1 & 0 & \ddots &  \\
\vdots & \vdots & \ddots & \ddots& J_{N-1}\\
0 & 0 &  & J_{N-1} & 0
\end{array}
\right).
\label{Ham-M}
\end{gather}
The dynamics (time evolution) of the system is completely determined by the eigenvalues $\epsilon_j$
and eigenvectors $\varphi_j$ of this interaction matrix.
It is indeed a standard technique~\cite{Lieb1968,Albanese2004}
to describe the $n$-fermion eigenstates of $\hat H$ ($n\leq N$) using the
single-fermion eigenstates $\varphi_j$ and Slater determinants.
In Christandl's case, determined by~\eqref{J}, the eigenvalues and eigenvectors of $M$ are explicitly known.
In particular, the eigenvalues are given by $\epsilon_j=-N+2j$ ($j=0,1,\ldots,N$) and the eigenvectors are given in terms of Krawtchouk polynomials.

In the model of Christandl there is, for arbitrary $N$, an analytic (closed form) expression for the eigenvalues and eigenvectors of $M$.
Such spin chains are {\em analytically solvable}~\cite{Chakrabarti2010,Jafarov2010}.
It implies, in particular, that the correlation function at time $t$,
\begin{gather*}
f_{r,s}(t) = (r| \exp(-it\hat H) |s),
\end{gather*}
where $r$ and $s$ are site labels belonging to $\{0,1,\ldots,N\}$ and $|r)$, $|s)$ denote the corresponding single spin states at the `receiver' and
`sender' sites $r$ and $s$,
can be computed explicitly~\cite{Chakrabarti2010,Jafarov2010}.

Christandl's spin chain model allows {\em perfect state transfer}, essential for using the spin chain as a transmission channel.
Perfect state transfer at time $t=T$ from one end of the chain to the other end is expressed by $|f_{N,0}(T)|=1$.
The topic of perfect state transfer in spin chains has received a lot of attention~\cite{Kay2010}.
Fairly easy suf\/f\/icient conditions have been formulated in order to achieve perfect state transfer (such as mirror symmetry~\cite{Albanese2004,Kay2010}).
Shi  et al.\ showed that the ``spectrum parity matching condition'' is necessary and suf\/f\/icient for perfect state transfer~\cite{Shi2005}.
According to this condition, they found a one-parameter extension of Christandl's model in the case there is an even number of
fermion sites in the chain; in our notation this means that $N$ is odd, i.e.
\begin{gather*}
N=2m+1, \qquad m\in{\mathbb Z}_+.
\end{gather*}
The couplings in Shi's model~\cite{Shi2005} are determined by ($k=0,1,\ldots,N-1$)
\begin{gather}
J_k= \begin{cases}
 \sqrt{(k+1)(N-k)}, & \text{if $k$ is odd,}\\
 \sqrt{(k+2\alpha+2)(N-k+2\alpha+1)}, & \text{if $k$ is even.}
 \end{cases}
\label{Ja}
\end{gather}
Herein, $\alpha$ is a real parameter satisfying $\alpha>-1$ (the case of Shi actually corresponds to half-integer $\alpha$, but for our purposes
$\alpha$ can be any real number greater than $-1$).
Note that for $\alpha=-\frac12$, Shi's model reduces to Christandl's model (i.e.~\eqref{Ja} reduces to~\eqref{J}), at least when $N=2m+1$ is odd.
In Shi et al.~\cite{Shi2005}, the spectrum of the single fermion states (i.e.\ the eigenvalues of $M$ with data determined by~\eqref{Ja}) was found;
however no closed form expressions for the eigenvectors were obtained.

In the present paper, we will show that the eigenvectors can be expressed in terms of Hahn polynomials.
In fact, we will f\/irst work with a two-parameter extension of Christandl's model:
\begin{gather}
J_k= \begin{cases}
 \sqrt{(k+1)(N-k)}, & \text{if $k$ is odd,}\\
 \sqrt{(k+2\alpha+2)(N-k+2\beta-1)}, & \text{if $k$ is even.}
 \end{cases}
\label{Jab}
\end{gather}
Now $\alpha$ and $\beta$ are real parameters satisfying $\alpha>-1$ and $\beta>0$.
The case of Shi corresponds to $\beta=\alpha+1$, and the case of Christandl to $\alpha=-\frac12$, $\beta=\frac12$.
For the general case~\eqref{Jab}, we obtain in this paper an explicit form of the eigenvalues, and an explicit form of the eigenvectors.
The components of the eigenvectors are given by means of Hahn polynomials $Q_n(x;\alpha,\beta,m)$~\cite{Koekoek,Suslov}: the even components are proportional to
$Q_n(x;\alpha,\beta,m)$ and the odd components to $Q_n(x; \alpha+1,\beta-1,m)$.
In order to prove our assertions about eigenvalues and eigenvectors, we need some (new) dif\/ference equations for Hahn polynomials.
Section~\ref{Hahn} of this paper is devoted to introducing the common notation for Hahn polynomials and to proving the new dif\/ference equations.
In Section~\ref{eigen} we obtain the main result of this paper: the explicit construction of the spectrum of $M$ for the values~\eqref{Jab} and the
construction of its eigenvectors in terms of the Hahn polynomials.
Section~\ref{correlation} returns to the model governed by the spin chain data~\eqref{Jab}. Since the spin chain is analytically solvable, we can compute the
correlation function explicitly, and determine under which conditions perfect state transfer is possible.
Finally, in Section~\ref{qHahn} we present the $q$-generalization of the results obtained (in terms of $q$-Hahn polynomials).

Although our paper is strongly inspired by the model introduced by Shi  et al., it should be emphasized that our results
are dealing mainly with mathematical aspects of this model.
In~\cite{Shi2005}, the emphasis was on quantum state transfer. Our main result is to show that the eigenstates of the Hamiltonian~\eqref{Ham2}
in the case of Shi, \eqref{Ja}, or in the extended case, \eqref{Jab}, can be computed in closed form, with coef\/f\/icients given as Hahn polynomial evaluations.

It should be mentioned that some completely dif\/ferent spin chain models related to Hahn polynomials have been considered before.
The second solution of~\cite{Albanese2004} is actually related to an interaction matrix corresponding to the Jacobi matrix of dual Hahn polynomials.
In~\cite{Chakrabarti2010}, the interaction matrix corresponding to the Jacobi matrix of Hahn polynomials was studied,
following some ideas of~\cite{Regniers2009}.
In that case, the matrix of eigenvectors $U$ is directly a matrix of Hahn polynomial evaluations.
However, due to the complicated coef\/f\/icients in the three term recurrence relations, the actual coef\/f\/icients in the interaction
matrix become quite involved, see e.g.~\cite[Lemma~2]{Chakrabarti2010}.
In the present paper, the main innovation comes from ``doubling'' the matrix~$U$ (hence the technique works for chains with an even length only),
in a way that it contains Hahn polynomial evaluations of two dif\/ferent types (one with parameters $(\alpha,\beta)$ and one with $(\alpha+1,\beta-1)$),
such that the interaction matrix coef\/f\/icients~\eqref{Jab} are very simple.

\section{Hahn polynomials and new dif\/ference equations}
\label{Hahn}

The Hahn polynomial $Q_n(x;\alpha, \beta, m)$~\cite{Koekoek,Suslov} of degree $n$ ($n=0,1,\ldots,m$) in the variable $x$, with parameters
$\alpha>-1$ and $\beta>-1$, or $\alpha<-m$ and $\beta<-m$
is def\/ined by~\cite{Koekoek,Suslov}:
\begin{gather}
Q_n(x;\alpha,\beta,m) = {\;}_3F_2 \left( \atop{-n,n+\alpha+\beta+1,-x}{\alpha+1,-m} ; 1 \right).
\label{defQ}
\end{gather}
Herein, the function $_3F_2$ is the generalized hypergeometric series~\cite{Bailey,Slater}:
\begin{gather}
{}_3F_2 \left( \atop{a,b,c}{d,e} ; z \right)=\sum_{k=0}^\infty \frac{(a)_k(b)_k(c)_k}{(d)_k(e)_k}\frac{z^k}{k!}.
\label{defF}
\end{gather}
In~(\ref{defQ}),  the series is terminating
because of the appearance of the negative integer $-n$ as a nume\-ra\-tor parameter.
Note that in~(\ref{defF}) we use  the common notation for  Pochhammer symbols~\cite{Bailey,Slater}
$(a)_k=a(a+1)\cdots(a+k-1)$ for $k=1,2,\ldots$ and $(a)_0=1$.
Hahn polynomials  satisfy a (discrete) orthogonality relation~\cite{Koekoek}:
\begin{gather}
\sum_{x=0}^m w(x;\alpha, \beta,m) Q_l(x;\alpha, \beta, m) Q_n(x;\alpha,\beta,m) = h_n(\alpha,\beta,m)  \delta_{ln},
\label{orth-Q}
\end{gather}
where
\begin{gather*}
  w(x;\alpha, \beta,m) = \binom{\alpha+x}{x} \binom{m+\beta-x}{m-x}, \qquad  x=0,1,\ldots,m, \\
  h_n (\alpha,\beta,m)= \frac{(-1)^n(n+\alpha+\beta+1)_{m+1}(\beta+1)_n n!}{(2n+\alpha+\beta+1)(\alpha+1)_n(-m)_n m!}.
\end{gather*}
Denote the orthonormal Hahn functions as follows:
\begin{gather}
\tilde Q_n(x;\alpha,\beta,m) \equiv \frac{\sqrt{w(x;\alpha,\beta,m)}\, Q_n(x;\alpha,\beta,m)}{\sqrt{h_n(\alpha,\beta,m)}}.
\label{Q-tilde}
\end{gather}

For our construction, the essential ingredient is a set of new dif\/ference equations for Hahn polynomials.
These relations involve Hahn polynomials of the same degree in variables $x$ or $x+1$, and with parameters $(\alpha,\beta)$ and $(\alpha+1,\beta-1)$;
in this sense it could also be appropriate to speak of ``contiguous relations'' rather than dif\/ference equations.
\begin{proposition}\label{proposition1}
The Hahn polynomials satisfy the following difference equations:
\begin{gather}
  (m+\beta-x) Q_n(x;\alpha ,\beta ,m)-(m-x) Q_{n}(x+1;\alpha , \beta,m)\nonumber \\
 \qquad{} =\frac{(n+\alpha +1)(n+\beta)}{\alpha+1} Q_n(x;\alpha+1,\beta-1,m), \label{Q-rec1} \\
  (x+1) Q_n(x;\alpha +1,\beta -1,m)-(\alpha +x+2) Q_{n}(x+1;\alpha +1 , \beta -1,m) \nonumber \\
 \qquad {}=-(\alpha+1) Q_n(x+1;\alpha,\beta,m). \label{Q-rec2}
\end{gather}
\end{proposition}

\begin{proof}
Both equations follow from a simple computation using the hypergeometric series expression.
In the case of~\eqref{Q-rec1}, the left hand side is expanded as follows:
\begin{gather*}
  (m+\beta-x) Q_n(x;\alpha ,\beta ,m)-(m-x) Q_{n}(x+1;\alpha , \beta,m) \nonumber\\
\qquad{} = \sum_{k=0}^n \frac{(-n)_k(\alpha+\beta+n+1)_k(-x)_{k-1}}{k!(\alpha+1)_k(-m)_k}
 [(\beta+m-x)(k-x-1)-(m-x)(-x-1)]. \nonumber
\end{gather*}
Rewriting the expression in square brackets as $[k(m-k+1)+(\beta+k)(k-x-1)]$, the above sum splits in two parts:
\begin{gather}
-\sum_{k=1}^n \frac{(-n)_k(\alpha+\beta+n+1)_k(-x)_{k-1}}{(k-1)!(\alpha+1)_k(-m)_{k-1}} +
\sum_{k=0}^n \frac{(-n)_k(\alpha+\beta+n+1)_k(-x)_{k}}{k!(\alpha+1)_k(-m)_k} (\beta+k).
\label{2parts}
\end{gather}
The f\/irst part can be brought in the following form:
\begin{gather*}
  -\sum_{k=1}^n \frac{(-n)_k(\alpha+\beta+n+1)_k(-x)_{k-1}}{(k-1)!(\alpha+1)_k(-m)_{k-1}}
  = -\sum_{j=0}^{n-1} \frac{(-n)_{j+1}(\alpha+\beta+n+1)_{j+1}(-x)_{j}}{j!(\alpha+1)_{j+1}(-m)_{j}} \\
\qquad{} = \sum_{k=0}^n \frac{(-n)_k(\alpha+\beta+n+1)_k(-x)_k}{k!(\alpha+1)_k (-m)_k} \frac{(n-k)(\alpha+\beta+n+k+1)}{(\alpha+k+1)}.
\end{gather*}
So \eqref{2parts} becomes
\begin{gather*}
  \sum_{k=0}^n \frac{(-n)_k(\alpha+\beta+n+1)_k(-x)_k}{k!(\alpha+1)_k (-m)_k}
\Bigl[ \frac{(n-k)(\alpha+\beta+n+k+1)}{(\alpha+k+1)} + (\beta+k) \Bigr] \nonumber\\
 \qquad{} =\sum_{k=0}^n \frac{(-n)_k(\alpha+\beta+n+1)_k(-x)_k}{k!(\alpha+1)_k (-m)_k}
\Bigl[ \frac{(n+\alpha+1)(n+\beta)}{(\alpha+k+1)} \Bigr] \nonumber\\
 \qquad{} = \frac{(n+\alpha+1)(n+\beta)}{(\alpha+1)} \sum_{k=0}^n \frac{(-n)_k(\alpha+\beta+n+1)_k(-x)_k}{k!(\alpha+2)_k (-m)_k} \nonumber
\end{gather*}
leading to the right hand side of~\eqref{Q-rec1}.

The second equation is even simpler to prove. The left hand side of~\eqref{Q-rec2} can be written as
\begin{gather}
(x+1)\sum_{k=0}^n A_k \frac{(-x+k-1)}{(\alpha+k+1)} - (\alpha+x+2) \sum_{k=0}^n A_k \frac{(-x-1)}{(\alpha+k+1)},
\label{prf2}
\end{gather}
where
\[
A_k= \frac{(-n)_k(\alpha+\beta+n+1)_k (-x)_{k-1}}{k!(\alpha+2)_{k-1} (-m)_k}.
\]
A simple addition of the two terms in~\eqref{prf2} yields
\[
- \sum_{k=0}^n A_k (-x-1) = -(\alpha+1) \sum_{k=0}^{n} \frac{(-n)_k(\alpha+\beta+n+1)_k (-x-1)_{k}}{k!(\alpha+1)_{k} (-m)_k},
\]
giving the right hand side of~\eqref{Q-rec2}.
\end{proof}

The set of dif\/ference equations will turn out to be the essential ingredient for the eigenvector construction in the next section.

\section{Eigenvalues and eigenvectors of the interaction matrix}
\label{eigen}

Let $N=2m+1$ be an odd integer, and consider the $(N+1)\times(N+1)$ interaction matrix $M$ of the form~\eqref{Ham-M} with
spin chain data $J_k$ determined by~\eqref{Jab}, i.e.\
\begin{gather}
J_k= \begin{cases}
 \sqrt{(k+1)(2m+1-k)}, & \text{if $k$ is odd,}\\
 \sqrt{(k+2\alpha+2)(2m+2\beta-k)}, & \text{if $k$ is even.}
 \end{cases}
\label{Jabm}
\end{gather}

We begin with the construction of a $(N+1)\times(N+1)$ matrix $U=(U_{ij})_{0\leq i,j \leq N}$ whose even rows
are given in terms of normalized Hahn polynomials~\eqref{Q-tilde} with parameters $(\alpha,\beta)$ and whose odd rows
are given in terms of normalized Hahn polynomials with parameters $(\alpha+1,\beta-1)$.
In order to have positive weight functions for both, we require that $\alpha>-1$ and $\beta>0$.
\begin{definition}
The $(N+1)\times(N+1)$ matrix $U$ with indices running from 0 to $N=2m+1$ is def\/ined by
\begin{gather}
  U_{2i,m-j} = U_{2i,m+j+1} = \frac{(-1)^i}{\sqrt{2}} \tilde Q_j(i;\alpha,\beta,m), \label{Ueven}\\
  U_{2i+1,m-j} = -U_{2i+1,m+j+1} = -\frac{(-1)^i}{\sqrt{2}} \tilde Q_j(i;\alpha+1,\beta-1,m), \label{Uodd}
\end{gather}
where $i,j\in\{0,1,\ldots,m\}$.
\label{defU}
\end{definition}

It is easy to verify that $U$ is an orthogonal matrix. Indeed, let us compute $U^TU$:
\begin{gather}
(U^TU)_{jk} = \sum_{i=0}^{2m+1} U_{ij}U_{ik}
= \sum_{i=0}^{m} U_{2i,j}U_{2i,k} + \sum_{i=0}^{m} U_{2i+1,j}U_{2i+1,k}.
\label{UU}
\end{gather}
For $j,k\in\{0,\ldots,m\}$, \eqref{UU} gives
\begin{gather*}
 \sum_{i=0}^{m} \frac{1}{2} \tilde Q_{m-j}(i;\alpha,\beta,m) \tilde Q_{m-k}(i;\alpha,\beta,m) \\
 \qquad \quad{}+ \sum_{i=0}^{m} \frac{1}{2} \tilde Q_{m-j}(i;\alpha+1,\beta-1,m) \tilde Q_{m-k}(i;\alpha+1,\beta-1,m) \\
\qquad{} = \frac{1}{2}  \delta_{m-j,m-k} + \frac{1}{2}  \delta_{m-j,m-k} =  \delta_{jk}
\end{gather*}
using the orthogonality~\eqref{orth-Q} of Hahn polynomials.
For $j,k\in\{m+1,\ldots,2m+1\}$, the computation is essentially the same and \eqref{UU} gives again~$\delta_{jk}$.
For $j\in\{0,\ldots,m\}$ and $k\in\{m+1,\ldots,2m+1\}$, \eqref{UU} gives
\begin{gather*}
 \sum_{i=0}^{m} \frac{1}{2} \tilde Q_{m-j}(i;\alpha,\beta,m) \tilde Q_{k-m-1}(i;\alpha,\beta,m) \\
 \qquad \quad{}- \sum_{i=0}^{m} \frac{1}{2} \tilde Q_{m-j}(i;\alpha+1,\beta-1,m) \tilde Q_{k-m-1}(i;\alpha+1,\beta-1,m) \\
\qquad{} = \frac{1}{2}  \delta_{m-j,k-m-1} - \frac{1}{2}  \delta_{m-j,k-m-1} =  0,
\end{gather*}
and for $j\in\{m+1,\ldots,2m+1\}$ and $k\in\{0,\ldots,m\}$, the result is the same.
So it follows that $(U^TU)_{jk}=\delta_{jk}$, or $U^TU=I$, the identity matrix. Hence $U^T$ is the inverse of $U$, so
$UU^T=I$ holds as well.

Now we have the main proposition.

\begin{proposition}\label{proposition2}
Let $M$ be the tridiagonal $(2m+2)\times(2m+2)$-matrix
\begin{gather*}
M= \left( \begin{array}{ccccc}
             0 & J_0  &    0   &        &      \\
            J_0 &  0  &  J_1  & \ddots &      \\
              0  & J_1  &   0  & \ddots &  0   \\
                 &\ddots & \ddots & \ddots & J_{2m} \\
                 &       &    0   &  J_{2m}  &  0
          \end{array} \right),
\end{gather*}
where the $J_k$ are given in~\eqref{Jabm}, and let $U$ be the matrix determined in Def\/inition~{\rm \ref{defU}}.
Then $U$ is an orthogonal matrix:
\begin{gather*}
U U^T = U^TU=I.
\end{gather*}
Furthermore, the columns of $U$ are the eigenvectors of $M$, i.e.
\begin{gather}
M U = U D,
\label{MUUD}
\end{gather}
where $D$ is a diagonal matrix containing the eigenvalues $\epsilon_j$ of $M$:
\begin{gather}
  D= \mathop{\rm diag}\nolimits  (\epsilon_0,\epsilon_1,\ldots,\epsilon_{2m+1}), \qquad
  \epsilon_{m-k}=-2\sqrt{(\alpha+k+1)(\beta+k)},\nonumber\\
  \epsilon_{m+k+1}=2\sqrt{(\alpha+k+1)(\beta+k)},
  \qquad  k=0,1,\ldots,m.\label{epsilon}
\end{gather}
\end{proposition}

\begin{proof}
The orthogonality of $U$ has already been proved, so it remains to verify~\eqref{MUUD} and~\eqref{epsilon}.
Now
\begin{gather}
\big(MU\big)_{ij}= \sum_{k=0}^{2m+1}M_{ik}U_{kj}=J_{i-1}U_{i-1,j}+J_{i}U_{i+1,j}.
\label{MU}
\end{gather}
We have to consider \eqref{MU} in four distinct cases, according to $i$ even or odd, and to $j$ belonging to $\{0,1,\ldots,m\}$ or to
$\{m+1,m+2,\ldots,2m+1\}$.
Let us consider the case that $i$ is odd and $j\in\{0,1,\ldots,m\}$. Then, relabelling the indices appropriately, \eqref{Jabm} and~\eqref{Ueven} yield:
\begin{gather*}
 (MU)_{2i+1,m-j}=J_{2i}U_{2i,m-j}+J_{2i+1}U_{2i+2,m-j}  \\
\phantom{(MU)_{2i+1,m-j}}{}  =(-1)^i\sqrt{2}\sqrt{(\alpha+i+1)(m+\beta-i)}\tilde Q_j(i;\alpha,\beta,m) \\
\phantom{(MU)_{2i+1,m-j}=}{} + (-1)^{i+1}\sqrt{2}\sqrt{(i+1)(m-i)}\tilde Q_j(i+1;\alpha,\beta,m)\\
\phantom{(MU)_{2i+1,m-j}}{}= (-1)^i \sqrt{2}\sqrt{\frac{(\alpha+1)_{i+1}(\beta+1)_{m-i-1}}{i!(m-i)!h_j(\alpha,\beta,m)}}[ (\beta+m-i) Q_j(i;\alpha,\beta,m)\\
\phantom{(MU)_{2i+1,m-j}=}{}
 -(m-i) Q_j(i+1;\alpha,\beta,m)].
\end{gather*}
Applying~(\ref{Q-rec1}), this becomes
\begin{gather*}
  =(-1)^i \sqrt{2}  \sqrt{\frac{(\alpha+1)_{i+1}(\beta+1)_{m-i-1}}{i!(m-i)!h_j(\alpha,\beta,m)}} \frac{(\alpha+j+1)(\beta+j)}{(\alpha+1)} Q_j(i;\alpha+1,\beta-1,m)\\
 = -2\sqrt{(\alpha+j+1)(\beta+j)} U_{2i+1,m-j} = \epsilon_{m-j}U_{2i+1,m-j}=\big(UD\big)_{2i+1,m-j}.
\end{gather*}
For $i$ odd and $j\in\{m+1,m+2,\ldots,2m+1\}$, the computation is essentially the same.
For $i$ even (and the two cases for~$j$), the computation is also similar, but now the second dif\/ference equation~\eqref{Q-rec2} must be used.
\end{proof}

Note that the spectrum of $M$ is symmetric, consisting of the values $\pm 2\sqrt{(\alpha+k+1)(\beta+k)}$ ($k=0,1,\ldots,m$).
Furthermore, when $\beta=\alpha+1$, the spectrum consists of integers $\pm 2(\alpha+k+1)$. This latter case
corresponds to the model of Shi  et al.~\cite{Shi2005}.

\section{Some aspects of the corresponding spin chain model}
\label{correlation}

Let us consider a spin chain~\eqref{Ham2} with data determined by~\eqref{Jab}.
The dynamics of this system is described by the unitary time evolution operator $\exp(-it\hat H)$.
The transition amplitude of a single spin excitation from site~$s$ to site $r$ of the
spin chain is given by the time-dependent correlation function~\cite{Bose2007,Chakrabarti2010}
\[
f_{r,s}(t) = (r| \exp(-it\hat H) |s).
\]
But the (orthonormal) eigenvectors of $\hat H$ in the single fermion mode are now known and given by
$\varphi_j= \sum\limits_{k=0}^N U_{kj}\,|\,k)$, i.e.\ the columns of the matrix $U$ constructed in~\eqref{Ueven}, \eqref{Uodd},
with $\hat H\varphi_j = M\varphi_j = \epsilon_j \varphi_j$.
Using the orthogonality of the states $\varphi_j$, one f\/inds~\cite{Chakrabarti2010}:
\begin{gather*}
f_{r,s}(t) =  \sum_{j=0}^N U_{rj}U_{sj}  {\rm e}^{-it\epsilon_j}.
\end{gather*}

Due to the expressions~\eqref{Ueven}, \eqref{Uodd}, implying $U_{r,m-j}=(-1)^r U_{r,m+j+1}$,
it is appropriate to write the correlation function in the following form:
\begin{gather}
f_{rs}(t) =\sum_{j=0}^m\big(U_{r,m-j}U_{s,m-j}  {\rm e}^{-it\epsilon_{m-j}}+
 U_{r,m+j+1}U_{s,m+j+1}  {\rm e}^{-it\epsilon_{m+j+1}}\big) \nonumber\\
\phantom{f_{rs}(t)}{} =\sum_{j=0}^m U_{r,m-j}U_{s,m-j} \big( \,{\rm e}^{-it\epsilon_{m-j}} +(-1)^{r+s} \,{\rm e}^{it\epsilon_{m-j}} \big).  \label{CF}
\end{gather}
Now it is a matter of considering the dif\/ferent parities for $r$ and $s$.
In the case they are both even, one f\/inds
\begin{gather*}
f_{2k,2l}(t)  =(-1)^{k+l}\sqrt{w(k;\alpha,\beta,m)w(l;\alpha,\beta,m)}\nonumber\\
 \phantom{f_{2k,2l}(t)  =}{} \times \sum_{j=0}^{m}Q_j(k;\alpha,\beta,m) Q_j(l;\alpha,\beta,m)
\frac{\cos\big(2t\sqrt{(\alpha+j+1)(\beta+j)} \big)}{h_j(\alpha,\beta,m)}. 
\end{gather*}
In the case the f\/irst index is odd and the second even, this becomes
\begin{gather}
f_{2k+1,2l}(t)  =-i(-1)^{k+l}\sqrt{w(k;\alpha+1,\beta-1,m)w(l;\alpha,\beta,m)}\label{CF2}\\
\phantom{f_{2k+1,2l}(t)  =}{} \times \sum_{j=0}^{m}Q_j(k;\alpha+1,\beta-1,m) Q_j(l;\alpha,\beta,m)
\frac{\sin\big(2t\sqrt{(\alpha+j+1)(\beta+j)} \big)}{\sqrt{h_j(\alpha+1,\beta-1,m)h_j(\alpha,\beta,m)}}. \nonumber
\end{gather}
The expressions for the case even/odd and odd/odd are similar, the main message being that due to the analytic
expressions for the eigenvectors, we obtain explicit expressions for the correlation function.

Let us examine, in this context, the transition from one end of the chain ($s=0$) to the f\/inal end of the chain ($r=N=2m+1$).
Expression~\eqref{CF2} reduces to:
\begin{gather}
f_{2m+1,0}(t)  = -i(-1)^m\sqrt{(\beta)_{m+1}(\alpha+1)_{m+1}} \nonumber\\
\phantom{f_{2m+1,0}(t)  =}{} \times \sum_{j=0}^m \frac{(2j+\alpha+\beta+1)(-m)_j}{(j+\alpha+\beta+1)_{m+1} j!}
\frac{\sin\big(2t\sqrt{(\alpha+j+1)(\beta+j)} \big)}{\sqrt{(\alpha+j+1)(\beta+j)}}.
\label{CF3}
\end{gather}
For general $\alpha$ and $\beta$, this expression cannot be simplif\/ied further.
Let us now consider the special case that
\[
\beta=\alpha+1.
\]
Then~\eqref{CF3} reduces to
\begin{gather}
f_{2m+1,0}(t) = -2i(-1)^m (\alpha+1)_{m+1} \sum_{j=0}^m \frac{(-m)_j}{(j+2\alpha+2)_{m+1} j!}
\sin\big(2t(\alpha+j+1)\big).
\label{CF3a}
\end{gather}
This can be rewritten as
\begin{gather}
f_{2m+1,0}(t) = -2i(-1)^m \frac{(\alpha+1)_{m+1}}{(2\alpha+2)_{m+1}} \sum_{j=0}^m  \frac{(-m)_j(2\alpha+2)_j}{j!(2\alpha+m+3)_j}
\sin\big(2t(\alpha+j+1)\big).
\label{CF4}
\end{gather}
The last sum is of hypergeometric type ${}_2F_1$, and so it can be further simplif\/ied for special values of $t$ and/or $\alpha$.
In particular, for $t=T=\pi/2$, one has
$\sin\big(\pi(\alpha+j+1)\big)=-(-1)^j \sin(\pi\alpha)$.
Using then Kummer's summation formula~\cite{Bailey,Slater}
\[
{}_2F_1 \left( \atop{-m,2\alpha+2}{2\alpha+m+3} ; -1 \right)= \frac{(2\alpha+3)_m}{(\alpha+2)_m},
\]
in the right hand side of~\eqref{CF4}, this expression reduces to
\begin{gather}
f_{N,0}\left(\frac{\pi}{2}\right)= f_{2m+1,0}\left(\frac{\pi}{2}\right) = i(-1)^m\sin(\pi\alpha).
\label{CF5}
\end{gather}
Note, by the way, that also for $t=2T=\pi$ one can simplify~\eqref{CF4}, since now
$\sin\big(2\pi(\alpha+j+1)\big)=\sin(2\pi\alpha)$.
Then, using Gauss's summation formula~\cite{Bailey,Slater}
\[
{}_2F_1 \left( \atop{-m,2\alpha+2}{2\alpha+m+3} ; 1 \right)= \frac{(m+1)_m}{(2\alpha+m+3)_m},
\]
the right hand side of~\eqref{CF4} reduces to
\begin{gather*}
f_{N,0}(\pi)= f_{2m+1,0}(\pi) = -2i\sin(2\pi\alpha) (-1)^m \frac{(\alpha+1)_{m+1}(m+1)_m}{(2\alpha+2)_{2m+1}} .
\end{gather*}

Note the importance of \eqref{CF5}. Indeed, keeping in mind that $\alpha>-1$, we have
\begin{gather}
\left|f_{N,0}\left(\frac{\pi}{2}\right)\right|= \left|f_{2m+1,0}\left(\frac{\pi}{2}\right)\right| = 1 \qquad \hbox{for} \quad
\alpha=-\frac12, \frac12, \frac32, \frac52,\ldots.
\label{psf}
\end{gather}
So there is perfect state transfer in the chain for $\alpha$ assuming one of these values, at time $t=\pi/2$.
In fact, this corresponds to the values given by Shi  et al.~\cite{Shi2005}.
Note that for $\alpha=-\frac12$, the spin chain data further reduces to that of Christandl~\cite{Christandl2004}.

The case~\eqref{psf} corresponds to $2\alpha+1 =2l$ with $l$ a nonnegative integer. As a matter of fact, this can
still be extended slightly. Let $2\alpha+1 =\frac{2l}{2k+1}$ with both $l$ and $k$ nonnegative integers.
Then for $t=T'=(2k+1)\pi/2$, the factor in~\eqref{CF4} becomes $\sin(2t(\alpha+j+1))=-(-1)^j \sin((2k+1)\alpha\pi)=(-1)^{j+k+l}$,
and so we have the result
\begin{gather*}
\left|f_{N,0}\left((2k+1)\frac{\pi}{2}\right)\right|= \left|f_{2m+1,0}\left((2k+1)\frac{\pi}{2}\right)\right| = 1 \qquad \hbox{for} \quad
2\alpha+1=\frac{2l}{2k+1}, \quad  k,l\in{\mathbb Z}_+ .
\end{gather*}
This case appears already in the paper of Qian  et al.~\cite{Qian}, who use the ``mirror mode concurrence''
to f\/ind this extension of Shi's result.

As far as perfect state transfer is concerned, our extension of Shi's model by an extra para\-me\-ter $\beta$ does not
give rise to any new cases. In fact, just for investigating perfect state transfer, the mathematical machinery developed here
is not necessary: the verif\/ication of the spectrum parity matching condition, using the entries in the interaction matrix and
the spectrum itself, is suf\/f\/icient.
The main advantage of our analysis is the explicit computation of the correlation function.
In particular, the simplicity of the expressions~\eqref{CF3a} and~\eqref{CF5}, describing the transfer from
one end of the chain to the other end, is striking.
Apart from the model of Christandl~\cite{Christandl2004}, where the general correlation function is given in~\cite[\S~2]{Chakrabarti2010},
there are no other models with such an elegant and simple correlation function.

\section[On the $q$-generalization of the previous results]{On the $\boldsymbol{q}$-generalization of the previous results}
\label{qHahn}

As the classical orthogonal polynomials of hypergeometric type have a generalization in terms of basic hypergeometric series, i.e.\
a $q$-generalization, one may wonder whether the present construction of the tridiagonal interaction matrix $M$ can also be generalized.
This is indeed the case: we can present a matrix $M_q$, whose eigenvectors are given in terms of $q$-Hahn polynomials, and whose
eigenvalues are symmetric and take a simple form.
In order to present these results, let us f\/irst brief\/ly recall some notation related to $q$-series~\cite{Gasper}.

For a positive real number $q$ ($\ne 1$), the $q$-Hahn polynomial $Q_n(q^{-x};\alpha, \beta, m |q)$ of degree $n$ ($n=0,1,\ldots,m$) in $q^{-x}$
is def\/ined by~\cite{Koekoek,Gasper}:
\begin{gather}
Q_n(q^{-x};\alpha,\beta,m|q) = {}_3\Phi_2 \left( \atop{q^{-n},\alpha\beta q^{n+1},q^{-x}}{\alpha q,q^{-m}} ; q,q \right).
\label{defqQ}
\end{gather}
Herein, the function $_3\Phi_2$ is the basic hypergeometric series~\cite{Bailey,Slater,Gasper}:
\begin{gather}
 {\;}_3\Phi_2 \left( \atop{a,b,c}{d,e} ; q, z \right)=\sum_{k=0}^\infty \frac{(a, b, c; q)_k}{(q,d, e;q)_k}
 z^k.
\label{defPhi}
\end{gather}
Note that in~(\ref{defPhi}) we use  the common notation for $q$-shifted factorials and their products~\cite{Gasper}:
\begin{gather*}
 (a_1,a_2,\dots ,a_A;q)_k=(a_1;q)_k(a_2;q)_k\cdots(a_A;q)_k,\\ 
 (a;q)_k=(1-a)(1-aq)\cdots\big(1-aq^{k-1}\big) \qquad {\rm and} \qquad (a)_0=1.
\end{gather*}
In~(\ref{defqQ}),  the series is terminating
because of the appearance of  $q^{-n}$ in the  numerator.
$q$-Hahn polynomials  satisfy a (discrete)  orthogonality relation~\cite{Koekoek}:
\begin{gather*}
\sum_{x=0}^m w(x;\alpha, \beta,m|q) Q_l\big(q^{-x};\alpha, \beta, m|q\big) Q_n\big(q^{-x};\alpha,\beta,m|q\big) = h_n(\alpha,\beta,m|q)  \delta_{ln},
\end{gather*}
where
\begin{gather*}
  w(x;\alpha, \beta,m|q) = \frac{(\alpha q,q^{-m};q)_x}{(q,\beta^{-1}q^{-m};q)_x} (\alpha \beta q)^{-x}, \qquad  x=0,1,\ldots,m, \\
  h_n (\alpha,\beta,m|q)= \frac{(\alpha\beta q^2;q)_{m}(q,\alpha\beta q^{m+2},\beta q;q)_n
(1-\alpha\beta q)(-\alpha q)^n}{(\beta q;q)_m (\alpha q)^m (\alpha q, \alpha\beta q, q^{-m};q)_n(1-\alpha\beta q^{2n+1})}
q^{(\atop{n}{2}) -mn}.
\end{gather*}
We shall assume that $0<q<1$; then the weight function is positive when $0<\alpha<q^{-1}$ and $0<\beta<q^{-1}$.
Denote the orthonormal $q$-Hahn functions as follows:
\begin{gather}
\tilde Q_n(q^{-x};\alpha,\beta,m|q) \equiv \frac{\sqrt{w(x;\alpha, \beta,m|q)}  Q_n(q^{-x};\alpha,\beta,m|q)}{\sqrt{h_n(q;\alpha, \beta,m)}}.
\label{qQtilde}
\end{gather}

Just as in Section~\ref{Hahn}, the main result needed is a set of dif\/ference equations for $q$-Hahn polynomials.
\begin{proposition}\label{proposition3}
$q$-Hahn polynomials satisfy the following $q$-difference equations:
\begin{gather}
 \big(1-\beta q^{m-x}\big) Q_n\big(q^{-x};\alpha ,\beta ,m|q\big) -\big(1-q^{m-x}\big) Q_{n}\big(q^{-x-1};\alpha , \beta,m |q\big)\nonumber \\
 \qquad {} =\frac{(1-\alpha q^{n+1})(1-\beta q^n) q^{m-n-x}}{1-\alpha q} Q_n\big(q^{-x};\alpha q,\beta q^{-1},m|q\big), \label{qQ-rec1} \\
 \big(1-q^{x+1}\big) \alpha q Q_n\big(q^{-x};\alpha q,\beta q^{-1},m|q\big)-\big(1-\alpha q^{x+2}\big) Q_{n}\big(q^{-x-1};\alpha q , \beta q^{-1},m|q\big)  \nonumber \\
 \qquad {} =-(1-\alpha q) Q_n\big(q^{-x-1};\alpha,\beta,m|q\big). \label{qQ-rec2}
\end{gather}
\end{proposition}

\begin{proof}
The proof follows the same computation as in the proof of Proposition~\ref{proposition1},
with the replacement of Pochhammer symbols by the corresponding $q$-shifted factorials (and keeping track of the
appropriate powers of~$q$).
\end{proof}

We now come to the construction of the tridiagonal matrix~$M_q$ and the matrix of eigenvectors~$U$.
The polynomials that appear here will be $q$-Hahn polynomials with parameters $(\alpha,\beta)$ and
$(\alpha q,\beta q^{-1})$. So in order to have positive weight functions for both sets of polynomials,
we shall assume:
\[
0<q<1, \qquad 0<\alpha<q^{-1}, \qquad 0<\beta<1.
\]
As before, let $N=2m+1$, and consider the $(N+1)\times(N+1)$ interaction matrix $M_q$ of the form~\eqref{Ham-M} with
non-zero matrix elements given by:
\begin{gather}
J_{2k+1} = 2\sqrt{(1-q^{k+1})(1-q^{m-k})q^{k+1}\alpha },  \qquad
J_{2k} = 2\sqrt{(1-\alpha q^{k+1})(1-\beta q^{m-k})q^{k}} ,
\label{qJi}
\end{gather}
where $k=0,1,\ldots,m$.
The $(N+1)\times(N+1)$ matrix $U$ with indices running from 0 to $N=2m+1$ is def\/ined similarly as in Def\/inition~\ref{defU},
but in terms of $q$-Hahn polynomials~\eqref{qQtilde}:
\begin{gather}
  U_{2i,m-j} = U_{2i,m+j+1} = \frac{(-1)^i}{\sqrt{2}} \tilde Q_j\big(q^{-i};\alpha,\beta,m|q\big), \label{qUeven}\\
 U_{2i+1,m-j} = -U_{2i+1,m+j+1} = -\frac{(-1)^i}{\sqrt{2}} \tilde Q_j\big(q^{-i};\alpha q,\beta q^{-1},m|q\big), \label{qUodd}
\end{gather}
where $i,j\in\{0,1,\ldots,m\}$. By the same argument as in Section~\ref{eigen}, the orthogonality of the matrix $U$ follows from the orthogonality
of the $q$-Hahn polynomials and the appropriate signs.

Then the main result in the $q$-case reads:
\begin{proposition}
Let $M_q$ be the tridiagonal $(2m+2)\times(2m+2)$-matrix
\begin{gather*}
M_q= \left( \begin{array}{ccccc}
             0 & J_0  &    0   &        &      \\
            J_0 &  0  &  J_1  & \ddots &      \\
              0  & J_1  &   0  & \ddots &  0   \\
                 &\ddots & \ddots & \ddots & J_{2m} \\
                 &       &    0   &  J_{2m}  &  0
          \end{array} \right),
\end{gather*}
where the $J_k$ are given in~\eqref{qJi}, and let $U$ be the matrix determined in~\eqref{qUeven}, \eqref{qUodd}.
Then $U$ is an orthogonal matrix, $U U^T = U^TU=I$.
Furthermore, the columns of $U$ are the eigenvectors of~$M_q$, i.e.
\begin{gather*}
M_q U = U D,
\end{gather*}
where $D$ is a diagonal matrix containing the eigenvalues $\epsilon_j$ of $M_q$:
\begin{gather*}
  D= \mathop{\rm diag}\nolimits  (\epsilon_0,\epsilon_1,\ldots,\epsilon_{2m+1}), \qquad
  \epsilon_{m-k}=-2\sqrt{(1-\alpha q^{k+1})(1-\beta q^k)q^{m-k}}, \nonumber\\
   \epsilon_{m+k+1}=2\sqrt{(1-\alpha q^{k+1})(1-\beta q^k)q^{m-k}},
  \qquad  k=0,1,\ldots,m. 
\end{gather*}
\end{proposition}

The proof of this proposition is essentially the same as that of Proposition~\ref{proposition3}, and uses the $q$-dif\/ference equations \eqref{qQ-rec1}, \eqref{qQ-rec2}.
Note that, as in the ordinary case, the spectrum of~$M_q$ is symmetric.

Once the explicit eigenvectors of the Hamiltonian are known in the $q$-generalized case~\eqref{qJi},
one can compute the correlation function, using the same expression~\eqref{CF}.
The expressions become quite involved, so we give just one example here.
This is in the case of transition from one end ($s=0$) to the other end ($r=2m+1$) of the chain, and for $\beta =q \alpha$:
\begin{gather*}
f_{2m+1,0}(t)  =i(-1)^{m}q^{m/2} \alpha^{m/2}(\alpha q;q)_{m+1}\nonumber\\
\phantom{f_{2m+1,0}(t)  =}{} \times \sum_{j=0}^m (-1)^j\sin\big(2t\big(1-\alpha q^{j+1}\big)q^{(m-j)/2}\big)
q^{j^2/2}
\frac{(q^{m-j+1};q)_{j}(1+\alpha q^{j+1})}{(\alpha^2 q^{j+2};q)_{m+1}(q;q)_j}.
\end{gather*}
The $q$-generalization does not give rise to any special cases with perfect state transfer, however.

\section{Conclusions and outlook}

We have dealt with some mathematical aspects of a spin chain model of Shi  et al.~\cite{Shi2005}, which is a~one-parameter extension
of the popular spin chain introduced by Christandl et al.~\cite{Christandl2004}.
In Christandl's model, the single fermion eigenvalues and eigenvectors could easily be computed, the eigenvectors being related to Krawtchouk polynomials.
In Shi's model, with an extra parameter $\alpha$, there was so far no known expression of the eigenvectors.
In the current paper, we have shown that these eigenvectors can be expressed in terms of Hahn polynomials.
As a matter of fact, we f\/irst extend Shi's model by introducing an extra parame\-ter~$\beta$, and then construct the eigenvectors.
In this process two types of Hahn polynomials are involved, those with parameters $(\alpha,\beta)$ and those with $(\alpha+1,\beta-1)$.
These Hahn polynomials are appropriately combined in a matrix~$U$, yielding the eigenvectors wanted.
When $\beta=\alpha+1$, the two-parameter spin chain reduces to that of Shi.
And when $\alpha=-\frac12$, $\beta=\frac12$, the spin chain reduces to that of Christandl.
Note, by the way, that in this last case the eigenvectors (expressed in terms of Hahn polynomials) indeed reduce to the known ones (expressed
in terms of Krawtchouk polynomials).
This follows from the fact that when $\alpha=-\frac12$ and $\beta=\frac12$ the Hahn polynomials, which are ${}_3F_2$ series, reduce to ${}_2F_1$ series
according to
\begin{gather*}
{}_3F_2 \left( \atop{-s,s+1,-i}{1/2,-m} ; 1 \right)= (-1)^i \frac{\binom{2m+1}{2i}}{\binom{m}{i}} {\;}_2F_1 \left( \atop{-2i,-m-s-1}{-2m-1} ; 2 \right),\\
{}_3F_2 \left( \atop{-s,s+1,-i}{3/2,-m} ; 1 \right)= -\frac{(-1)^i}{(2s+1)} \frac{\binom{2m+1}{2i+1}}{\binom{m}{i}} {\;}_2F_1 \left( \atop{-2i-1,-m-s-1}{-2m-1} ; 2 \right).
\end{gather*}
These reductions can be obtained, e.g., from~\cite[(48)]{Atakishiyev2005}. The ${}_2F_1$ series in the right hand side correspond to (symmetric) Krawtchouk polynomials
(with $p=1/2$).

Due to the explicit forms of the eigenvectors, the time-dependent correlation function $f_{r,s}(t)$ has been computed for the spin chains under
consideration. In special cases, the expression of the correlation function is particularly simple, see e.g.~\eqref{CF5}.

In the construction of the eigenvectors, the main relations needed are two new dif\/ference equations for Hahn polynomials.
We have also examined the $q$-generalization of these results. The $q$-extension of these dif\/ference equations is more or less straightforward.
Also the construction of the corresponding eigenvectors in terms of $q$-Hahn polynomials has been completed.

The extension of symmetric Krawtchouk polynomials (without a parameter $\alpha$) to Hahn polynomials (with parameters $(\alpha,\alpha+1)$)
may also be used in other applications.
In particular, we hope to extend the f\/inite oscillator models of~\cite{Atakishiyev2005}, by introducing such an extra parameter.
It remains to be seen, in that case, how the underlying Lie algebra is deformed, and how the parameter has an inf\/luence on the
f\/inite oscillator eigenstates.
This topic will be treated elsewhere.

\subsection*{Acknowledgements}

N.I.~Stoilova would like to thank Professor H.D.~Doebner (Clausthal, Germany) for constructive discussions.
N.I.~Stoilova was supported by project P6/02 of the Interuniversity Attraction Poles Programme (Belgian State --
Belgian Science Policy) and by the Humboldt Foundation.

\pdfbookmark[1]{References}{ref}
\LastPageEnding

\end{document}